\documentclass[fleqn,usenatbib]{mnras}

\usepackage{newtxtext,newtxmath}

\usepackage[T1]{fontenc}

\DeclareRobustCommand{\VAN}[3]{#2}
\let\VANthebibliography\thebibliography
\def\thebibliography{\DeclareRobustCommand{\VAN}[3]{##3}\VANthebibliography}

\usepackage{graphicx}	
\usepackage{amsmath}	
\usepackage{xcolor}	

\title[Relics and cETGs: dynamics and environment]{The internal dynamics and environments of Relics and compact massive ETGs with TNG50} 

\author[Moura,~M.~T.~et al.]{
Micheli T. Moura,$^{1}$\thanks{E-mail: micheli\_trindade.moura@hotmail.com}
Ana L. Chies-Santos,$^{1}$\thanks{E-mail: ana.chies@ufrgs.br}
Cristina Furlanetto,$^{1}$\thanks{E-mail: cristina.furlanetto@ufrgs.br}
Ling Zhu$^{2}$, Marco A. Canossa-Gosteinski$^{1}$
\\
$^{1}$Instituto de Física, Universidade Federal do Rio Grande do Sul, Av. Bento Gonçalves 9500, Porto Alegre, RS. 90040-060, Brazil \\
$^{2}$Shanghai Astronomical Observatory, Chinese Academy of Sciences, 80 Nandan Road, Shanghai 200030, China\\
}

\date{Accepted 2023 December 25. Received 2023 December 23; in original form 2023 September 17
}

\pubyear{2015}

\begin{document}
\label{firstpage}
\pagerange{\pageref{firstpage}--\pageref{lastpage}}
\maketitle

\begin{abstract}
Relic galaxies are massive, compact, quiescent objects observed in the local Universe that have not experienced any significant interaction episodes or merger events since about $z = 2$, remaining relatively unaltered since their formation. On the other hand, massive and compact Early-type Galaxies (cETGs) in the local Universe appear to show similar properties to Relic galaxies, despite having substantial accretion history. Relic galaxies, with frozen history, can provide important clues about the intrinsic processes related to the evolutionary pathways of ETGs and the role that mergers play in their evolution. 
Using the high-resolution cosmological simulation TNG50-1 from the Illustris Project, we investigate the assembly history of a sample of massive, compact, old, and quiescent subhalos split by satellite accretion fraction. We compare the evolutionary pathways at three cosmic epochs: $z=2$, $z=1.5$, and $z = 0$, using the orbital decomposition numerical method to investigate the stellar dynamics of each galactic kinematical component and their environmental correlations. 
Our results point to a steady pathway across time that is not strongly dependent on the mergers or the environment. Relics and cETGs do not show a clear preference for high or low-density environments within the volume explored at $z=0$, as they are found in both scenarios. However, the progenitors of Relic galaxies have consistently resided in high-density environments since $z=2$, while cETGs were shifted to such environments at a later stage. The merger history can be recovered from the stellar kinematics imprints in the local Universe. Relics and cETGs show consistently dynamical similarities at $z=2$ and differences at $z=0$ to disk, bulge, and hot inner stellar halo. In the current scenario, the mergers that drive the growth of cETGs do not give rise to a new and distinct evolutionary pathway when compared to Relics.

\end{abstract}



\section{Introduction}

 It has been shown that the difference between massive early-type galaxies (ETGs) and their counterparts at high redshift ($z \gtrsim 2$) is relevant \citep{2005Daddi,2006ApJKriek_vanDokkum,2008Buitrago,2007Trujillo,van_der_Wel_2009,2009Cemarro&Trujillo,2014BelliandNewman}. Observations and numerical studies suggest that high-redshift ETGs tend to be compact and smaller in comparison with the local analogs, with half-light radii
smaller by a factor of $\sim 3$ \citep{vanderWel2014,2009Naab,2010VanDokkum,2010OserOstriker,2022Remus}.

A simple scenario for the formation and evolution of ETGs is the two-phase scenario \citep{2010OserOstriker}. The initial phase of formation at $z > 2$ is characterized by the rapid growth of the stellar mass by wet mergers leading to a phase with high star formation rate (SFR), while there is no significant growth of size (e.g., \citealt{2015ApJVanDokkun,2015Zolotov,2020Zibetti}). Due to the quick increase of mass (to about $10^{11} \rm{M_\odot}$) and star formation, one expects that these compact and massive objects quench and become passive. The first stage of formation is the so-called `red nugget'. The second stage refers to a growth in size surrounding the nugget by dry mergers with other smaller structures. The two-phase scenario associated with the assembly history of ETGs is currently being investigated by observational and numerical approaches \citep{2016CitroAnnalisa,2022Ditrani,Floresfreitas2022,2022Ji,2022ZhuLing,zhu2022b}. In this context, Relic galaxies are defined as massive, compact, and quiescent objects observed in the local Universe that have not experienced any significant interaction episodes or important mergers events since $z\sim 2$ (the preserved nugget), therefore remaining mainly unaltered since their formation. 

Relic galaxies can provide important clues about the stochastic processes related to the formation of massive ETGs \citep{2009Trujillo, 2017Ferré-Mateu,2022Salvador,2023Martin,2021Spiniello}. A common discussion topic is the impact of mergers and which are the dominant quenching mechanisms, and their relation with the morphological evolution of massive and passive galaxies found in the local Universe and across cosmic times \citep{Hopkins_2010,2015Zolotov,10.1093tacchella2016,2019Tacchella,2022Park,2022Ji}. 
 \citet{2016Davidzon} has investigated environmental effects in the evolution of ETGs galaxies at  $z \approx 0.5 -  0.9$, where the galaxy stellar mass function in the densest regions could have a different shape than what is measured at low densities, especially for massive galaxies. This result supports the progressive quenching due to internal processes in galaxies located in low dense environments \citep{2010Peng, Papovich2018}. At higher cosmic times, \citet{Taylor2023} have investigated the stellar mass functions for quiescent and post-starburst (PSB) galaxies, finding a larger number of passive galaxies in higher-density environments, compared to intermediate and low-density environments at $z=3$. Several mechanisms have been proposed to work together to quench galaxies. Some examples include internal processes such as gas removal from the AGN feedback processes \citep{2017Combes,2015Volonteri,2021Piotrowska}, morphological quenching \citep{1980Dressler,2011Cappelari}, and global environmental effects due to mergers, such as ram pressure stripping (RPS) and others \citep{1972Gunngott,2023Rohr,2019Vulcani}.

 Cosmological simulations have been proven to be an important tool to investigate the role of the environment in the evolutionary stages of galaxy evolution (e.g. \citealt{2012Gabor,2019Ruggiero,2023Hasan,2023Kulier}). As far as environmental effects are concerned, much remains to be investigated. \citet{2016PeraltaArriba} found a parallel between observations from NYU-VAGC catalog \citep{2005Blanton}, and simulations from modified Millenium-I run \citep{2005Springel}, reporting the Relics' preference for dense environments. Most recently, \citet{Floresfreitas2022} identified the most reliable Relics galaxies candidates in Illustris TNG50 simulation based on observational properties of such objects reported in the literature, in which one of the conclusions is that Relics at $z=0$ are closely connected to the environment in which their progenitors evolved.

Relic systems allow us to study the physical processes that shaped the mass assembly of massive galaxies in the early Universe and are currently reachable in the nearby Universe through high spatial resolution and high enough spectroscopic signal-to-noise ratio, achievable with current instrumentation. Using cosmological simulations, we aim to investigate the evolutionary assembly of massive compact ETGs along Relic galaxies, through the local dynamic on galactic scales, as well as understand how the environment might affect their evolution. This paper is organized to introduce the employed numerical methods and sample selection in Sect. \ref{Sec:Methods and data}. Results in Sect. \ref{Sec:results} are organized into two main branches, the internal dynamics and environment. Summary and discussion are in Sect. \ref{Sec:summary}. Throughout the paper, we assume the \citet{plank2016} parameters to $\Lambda$CDM cosmological model with $H_{0} = 67.74\,\rm{km}\,\rm{s}^{-1} \rm{Mpc}^{-1}, \Omega_{m} = 0.30$, and $\Omega_{\Lambda} = 0.69$.
\section{Methods and sample selection}
\label{Sec:Methods and data}

\subsection{TNG cosmological simulations and numerical model}
To investigate the hydrodynamic history of the simulated galaxies, this work utilizes the suite of IllustrisTNG cosmological magnetohydrodynamical simulations \citep{2018Nelson,2018Springel,2018Pillepich,2018Naiman,2018Marinacci,2021Nelson,2022Park}, using the moving-mesh \textsc{arepo} code \citep{2010Springel}. The TNG suite consists of three simulation boxes comprising different mass resolutions and physical sizes, allowing to explore manifold aspects, from clustering with the largest simulation box -- TNG300, to the structural properties of galaxies and their hydrodynamical inner processes with bigger mass resolution using TNG50 and TNG100. Aiming to reach the structural dynamic details about massive and compact galaxies, we use data from the TNG50 run, as it reaches the required numerical resolution to process the dynamical and hydrodynamical inner structures.

TNG50-1 run ($\rm{L}_{\rm box} = 51.7~\rm{cMpc}$) employs the \textsc{subfind} algorithm \citep{2001Springel,2009Dolag} to locate gravitationally bound haloes and subhalos structures, that comprise baryonic masses~$m_{\rm baryons}=8.5\times10^{4}\,\rm{M}_{\odot}$, and dark matter mass resolution of $m_{\rm DM}=4.5\times10^{5}\,\rm{M}_{\odot}$. The assembly history of the subhalos can be tracked through snapshots in time using \textsc{sublink} merger trees \citep{2015RodriguezG}, where a descending subhalo has one unique progenitor. Through the merger trees across the snapshots it is possible to follow the stellar particles defined as in-situ or ex-situ, as a result of mergers across cosmic times. TNG50 simulation recipes were improved to analyze internal kiloparsec-scale structure and dynamics, as described in \citet{2019Pillepich}, from $z \approx 0~\rm{to}~z\approx 6$, being this run (TNG50-1) suitable to investigate the dynamic history in small galactic scales over cosmic times.

Halos are identified from the particle distribution using a ‘friend-of-friends’ (\textsc{FoF}) algorithm, which connects together dark matter particles \citep{2001Davies}. Any particles within each \textsc{FoF} group that are gravitationally bound are identified using the \textsc{subfind} algorithm \citep{2001Springel,2009Dolag}, and defined as a subhalo, with the requirement that each subhalo-galaxy like contains at least 20 resolution mass elements \citep{2019Bose}. Every subhalo assigned as a member of a \textsc{FoF} group within $R_{200}$ belongs to their host halo. $R_{200}$ is defined as the radius within which the halo mean density is 200 times the critical density of the Universe at the given redshift.

\subsection{Sample selection}
\begin{figure}
\centering
\includegraphics[width=\linewidth]{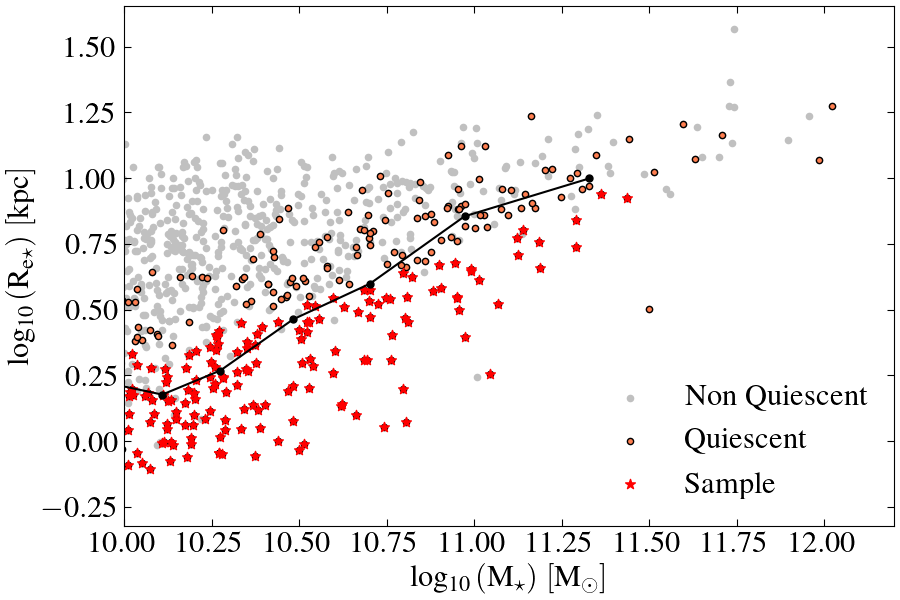}
\caption{Mass and size relation from Illustris TNG50 subhalos at $z=0$. gray dots are all non-quiescent subhalos, followed by the quiescent subhalos in orange, and the selected sample as red stars. Median values for the mass and size of the quiescent subhalos in each stellar mass bin are shown as the black solid line connected by the black dots.}
\label{Fig.1}
\end{figure}

We select a sample of 156 simulated compact galaxies with mass, size, quiescence, and age constraints of $\rm{(M_{\star})}\ge10^{10}\,\rm{M}_{\odot}$ effective radius $\rm({R}_{e})\le 4\,$kpc, specific Star Formation Rate $(\rm{sSFR})\,<10^{-11}\,\rm{M}_{\odot}\,\rm{yr^{-1}}$, and age $\ge 5\,$Gyr at $z=0$, to obtain the selected sample of massive, compact and quiescent galaxies shown as red stars in Fig.\,\ref{Fig.1}. The thresholds found in the literature for the compactness of massive quiescent ETGs vary from less to more restrictive criteria as shown in \citet{2023kristof}. Here we adopt the current mass-size relation to maximize the selected sample considering the TNG50 box size. 

ETGs in the local Universe differ in size from their counterparts at higher redshift ($z \gtrsim 2$) \citep{2005Daddi,2007Trujillo,2009Damjanov}. Relic galaxies at $z=0$ are red nuggets that survived dry mergers events without undergoing massive accretions since then. Consequently, massive compact ETGs and Relic galaxies appear to show similar properties in the local Universe despite their accretion history. In this paper we define the accretion based on the stellar mass of the satellite accretion fraction as: $\rm{M_{\star\,tot}}/M_{\star\,z=0}$, where $\rm{M_{\star\,tot}}$ is the total stellar mass of the accreated satellites. From the selected sample we consider an accretion cut of $10\%$ to separate the compact ETGs (cETGs) from the Relics subhalos.
The majority of massive subhalos in TNG50 have masses ranging from $10^{10} - 10^{10.5}\,\rm{M_\odot}$ (`low'-mass regime), while a smaller fraction falls into the supermassive subhalo regime with stellar masses exceeding $10^{10.5}\,\rm{M_\odot}$ (high-mass regime). The average accretion fraction for subhalos in these ranges varies from 7\% for the low-mass regime and increases to 40\% for masses around $10^{11}\,\rm{M_\odot}$. We defined a $10\%$ cutoff for the entire sample, as it proved suitable for both low and high-mass regimes. Otherwise, in the high-mass regime, this accretion fraction aligns with values obtained for the expected ex-situ accretion of NGC 1277 \citep{2018Beasley}. The final selected sample is composed of 156 old, massive, and compact subhalos at $z=0$, where through time 57 of them accreted more than $10\%$ of satellites (hereafter cETGs), while the remaining 99 subhalos stayed relatively `frozen' since old times (hereafter Relics). 

\begin{figure}
\centering
\includegraphics[width=1.0\linewidth]{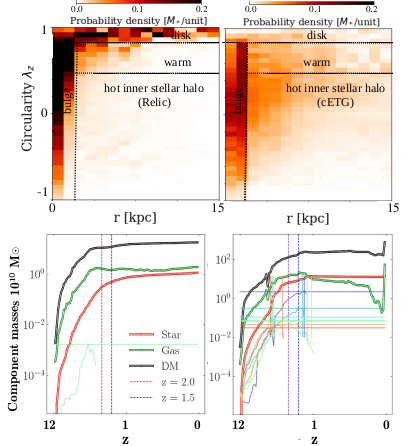}
\caption{Upper panels: Stellar orbit distribution, $p(r, \lambda_{z})$ in the phase-space of circularity $(\lambda_{z}, r)$ for two simulated galaxies in TNG50: Relic (left) and a cETG (right) at $z=0$. The kinematical components are separated by the dashed dark lines. Lower panels: Stellar, gas, and dark matter mass over time of the subhalos in the upper panels in red, green, and black lines, respectively. Dashed lines in magenta and blue indicate $z=2$ and $z=1.5$. Satellite accretion is illustrated as colored solid lines below the main components lines.}
\label{Fig.2}
\end{figure}

\subsection{Kinematical decomposition method}
\label{sub:orbitaldecomposition}

We perform a kinematical decomposition to separate different structures in a galaxy.
In the simulation context, the kinematic decomposition method has been investigated using different approaches, some examples include smoothing particle hydrodynamics (SPH) code \citep{2003Abadi}, zoom-in simulation \citep{2016MObreja}, implementing an unsupervised machine learning algorithm to decompose structures from cosmological simulations \citep{2019Du} and most recently, applying the technique into TNG50, TNG100, TNG300, and EAGLE cosmological simulation \citep{2022ZhuLing}.  
In these studies, the major parameters that characterize the stars on different structures are the energy and angular momentum.

The kinematic decomposition is also applied to the Milky Way with the 6D phase-space information of stars obtained from observations \citep[e.g.,][]{Helmi2018}. It has become possible for nearby galaxies by applying a \citet{1979Schwarzschild} orbit superposition dynamical method to galaxies observed by Integral Field Unit
(IFU) instruments. From the Schwarzschild model, the stellar orbit distribution of a galaxy could be obtained. Two parameters, circularity $\lambda_z$ and time-averaged radius $r$ are commonly used to characterize the stellar orbits, and the stellar orbit distribution is presented as the probability density distribution of the orbits in the phase space of $\lambda_z$ and $r$. Based on the stellar orbit distribution, kinematical decomposition can be performed on real galaxies in a similar way to the simulations.
This technique has been applied to a large sample of galaxies from the IFU spectroscopy surveys, such as SAURON \citep{2001Davies} by \citet{2008Vandebosch}, CALIFA \citep{2014WalcherCalifa} by \citet{2018ZhuLing}, SAMI \citep{Bryant2015} by \citet{Santucci2022}, MaNGA \citep{2016Bundy} by \citet{2020MNRASJinZhuLong}, and MUSE observed galaxies \citep{Ding2023} to investigate dynamical structures.

To make the structure decomposition comparable to that of real galaxies based on the orbit superposition method, in this paper, we also characterize the stellar orbit distribution of each simulated galaxy by the probability density distribution of $p(\lambda_z, r)$, as illustrated by two galaxies in Fig~\ref{Fig.2}. Then we perform the structure decomposition generally following \citet{2022ZhuLing}. We decompose each galaxy into four components: the disk $\lambda_{z}> 0.8$ and $r < r_{\rm{max}}$, the bulge $\lambda_{z} > 0.8$ and $r< r_{\rm{cut}}$, and the hot inner stellar halo (named simply as halo through the paper) component $\lambda_{z}< 0.5$ and $r_{\rm{cut}} < r < r_{\rm{max}}$, where $r_{\rm{cut}} = 2.5$ kpc and $r_{\rm{max}} = 7$ kpc, as shown on Fig.~\ref{Fig.2}. The four components are defined in a similar way as that in \citet{2022ZhuLing}, except that we choose $r_{\rm{cut}} = 2.5\,\rm{kpc}$, which showed more appropriate for the compact galaxies focused in this paper.

\vspace{-0.7cm}
\section{Results}
\label{Sec:results}
The results are organized into two main sections: the internal stellar dynamic of the sample, analyzed from the orbital decomposition method in Sect. \ref{Sec.Dynamics}, and Sect. \ref{Sec:environment}, where we discuss how the environment acts on Relics and cETGs.

\subsection{Internal dynamics}
\label{Sec.Dynamics}
In the following, we apply the numerical decomposition method described in Sect.~\ref{sub:orbitaldecomposition}, and present the results obtained for the mass evolution, ages, and metallicities of the sample.

\subsubsection{Mass evolution}
\label{sub:mass_evolution}

We investigate the sample outlined in three different cosmic epochs, $z=2$, $z=1.5$, and $z=0$, to understand the stellar mass evolution over each kinematical component named disk, bulge, and halo. Those specific redshifts were chosen to reach the dynamic settings of Relics at different stages, such as the `red nugget' stage at $z=2$ \citep{2020tortora,2021Spiniello,2023spiniello}. At $z=1.5$ most ETGs and Relics from our sample have passed through their more massive mergers, evolving mostly passively after this period, especially Relics. \citet{Floresfreitas2022} selected Relic candidates considering their passive evolution from this $z$ onward. The evolution ends at $z=0$, and can be compared to available and future observations of Relics and cETGs of the local Universe. 

\begin{figure}
\centering
\includegraphics[width=\linewidth]{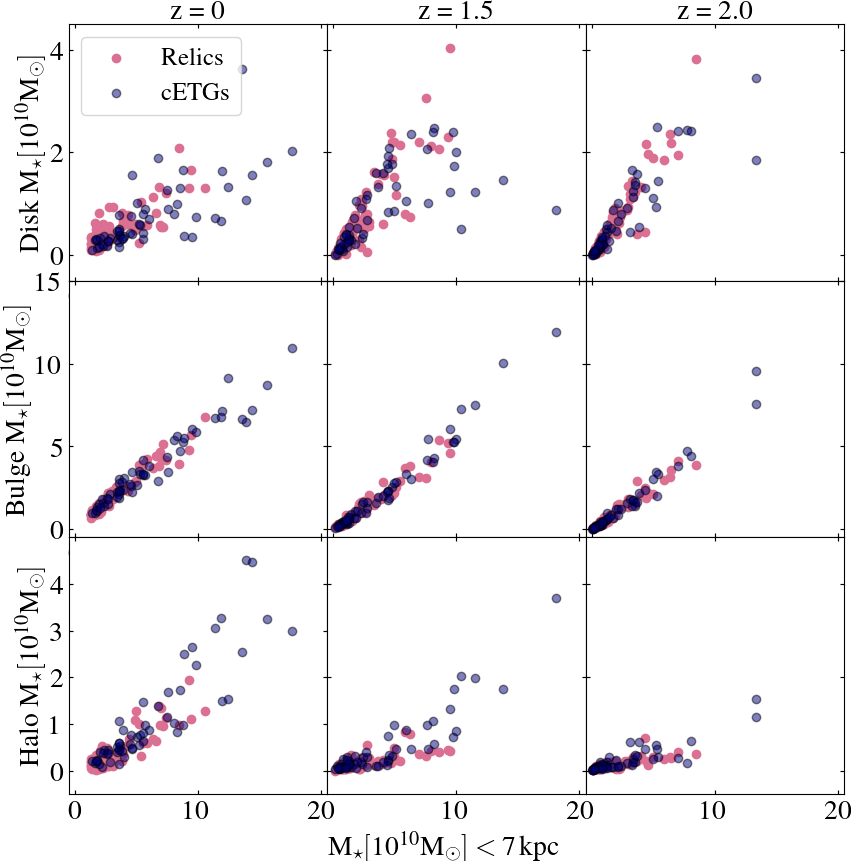}
\caption{Halo, Bulge, and Disk Stellar mass over total stellar mass within 7 kpc from the center of the subhalos, for cETGs (blue) and Relics (magenta). Each column represents a different redshift ($z=0,z=1.5,z=2$).}
\label{Fig.3}
\end{figure}

Fig.~\ref{Fig.3} presents the results for the stellar mass evolution of each kinematical component over time on the redshifts evaluated as a function of the total stellar mass of the subhalos within 7 kpc ($r_{\rm{max}}$). Each column shows the stellar mass for Relics and cETGs at the given time for the disk, bulge, and halo, at $z=0$ on the left, $z=1.5$ in the center, and $z=2$ in the right column. While Relics and cETGs consistently exhibit similar overall behavior across all redshifts considered, there is notable variation in the behavior of individual components when examined separately. To quantify the significance of the similarities and differences between the subsample distribution, we performed the Kolmogorov–Smirnov test (\textit{KS} - test) and Anderson-Darling test (\textit{AD} - test) \citep{1952AD}, through the kinematical components over each redshift. These tests assess whether two distributions could have come from similar underlying functions, evaluating the current data distribution. The \textit{KS}-test quantifies the maximum vertical distance between the cumulative distribution functions (CDFs) of two samples, while \textit{AD}-test assigns more weight to the tails of the distributions, making it sensitive at detecting deviations in the extremes. The results for \textit{KS} and \textit{AD}-test shown in Tab.~\ref{Tab:table} agree in most cases, with the exception of the disk, bulge, and Fdisk mass at $z=1.5$, and halo mass at $z=2$, where the divergence between the tests exceeds $5\%$ in the analysis of the distribution. As result, Relics and cETGs present non-negligible similarities in terms of the overall distribution, although exhibit also kinematical differences in all evaluated components at $z=0$, without any divergence between the tests. Disk and bulge at $z=2$ present statistical similarities, implying that the stellar dynamics of these components were sufficiently similar at that epoch.
About $z=1.5$ the hot inner stellar halo already shows differences between the sample, although disk and bulge dynamics are inconclusive by then. This can indicate that at this moment a transition may be occurring toward the complete distinction of these components at $z=0$ for both subsamples.

As we have access to each kinematic component individually, we explored the stellar mass fraction of the disk over the stellar mass fraction of the halo of the sample, to evaluate the consistency of the previous result considering the disk and halo fractions over each time, as shown in Fig.~\ref{Fig.4}. This figure shows the behavior of the stellar disk mass fraction (Fdisk) vs. stellar halo mass fraction (Fhalo), for $z=0, z=1.5$, and $z=2$ in each column. The upper panels show both subsamples in the same frame, and the second and third rows show the isolated distribution for cETG and Relics separately. The significance of the similarities and differences of each kinematical fraction over time is indicated in Table.~\ref{Tab:table}. Considering each redshift, the results point out the same dynamical trends as in Fig.~\ref{Fig.3}, where at $z=0$ both Fdisk and Fhalo components are different among the sample, although they were dynamically comparable at $z=2$. At $z=1.5$ Fhalo appears different for cETGs and Relics, while the disk is still inconclusive at this point, as indicated in Fig.~\ref{Fig.3} as well. Those dynamical differences between Relics and cETGs can be attributed to the extended accretion history of cETGs, which might be reflected at $z=0$ in all components, and in early times in the hot inner stellar halo component. This component has shown evidence of ancient merger signatures embedded in the stellar dynamics persisting until $z=0$.
The stellar kinematics properties shown may point to a predominance of local processes to their evolution (i.e. internal stellar dynamics) instead of the global processes in our compact galaxies. As we will see in the next sections, despite the environmental differences and the cETGs extended accretion, the homogeneity of other properties such as ages, quiescence, metallicity, and compactness among the sample are maintained through time.

\begin{table}
    \centering
    \caption{Kolmogorov–Smirnov statistical test (\textit{KS} - test) and Anderson-Darling test (\textit{AD} - test) to each kinematical component at the given redshift (\textit{z}) for Relics and cETGs from Fig.~\ref{Fig.3} and Fig.~\ref{Fig.4}.}
    \label{Tab:table}
    \begin{tabular}{lcccr}
        \hline
        Component & \textit{z} & KS & AD\\
        \hline
        Disk      & 0   & 0.00595 & 0.00281\\
        Disk      & 1.5 & 0.00828 & 0.08424\\
        Disk      & 2.0 & 0.48474 & 0.25\\
        Bulge     & 0   & 3.93$\rm{e}^{-08}$ & 0.001\\
        Bulge     & 1.5 & 0.03422 & 0.00630\\
        Bulge     & 2.0 & 0.46020 & 0.17144\\
        Halo      & 0   & 2.72$\rm{e}^{-10}$ & 0.001\\
        Halo      & 1.5 & 0.00038 & 0.001\\
        Halo      & 2.0 & 0.10524 & 0.02306\\
        Fdisk     & 0   & 0.00266 & 0.00112\\
        Fdisk     & 1.5 & 0.01398 & 0.00757\\
        Fdisk     & 2.0 & 0.43123 & 0.25\\
        Fhalo     & 0 &  3.29$\rm{e}^{-08}$ & 0.001\\
        Fhalo     & 1.5 & 0.00828 & 0.00170\\
        Fhalo     & 2.0 &  0.58307 & 0.25\\

        \hline
    \end{tabular}
\end{table}

 \subsubsection{Mass, age, and metallicity}

\begin{figure}
\centering
\includegraphics[width=\linewidth]{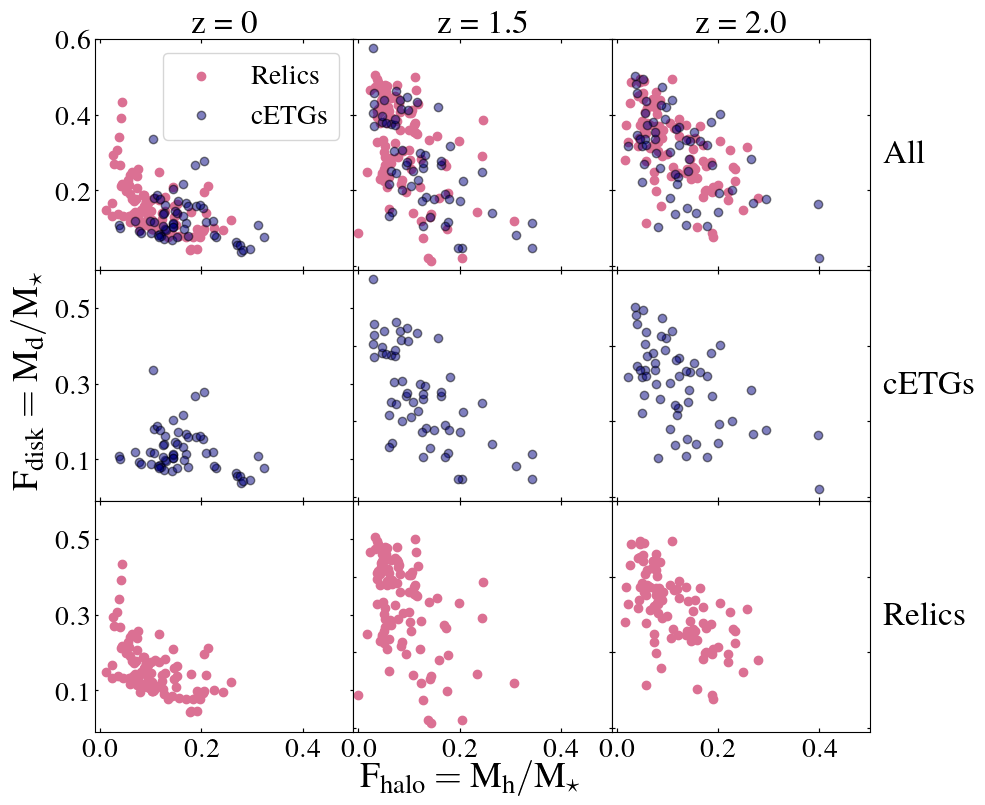}
\caption{Evolution of the stellar disk and stellar halo mass fractions. Each column shows a redshift: $z =0$, $z=1.5$, and $z = 2.0$ from left to right. The first line (from top to bottom) shows both the Relics in magenta and cETGs in blue. The second and third lines show only the cETGs stellar mass fractions in blue and Relics in magenta, respectively.}
\label{Fig.4}
\end{figure}

\begin{figure}
\centering
\includegraphics[width=\columnwidth]{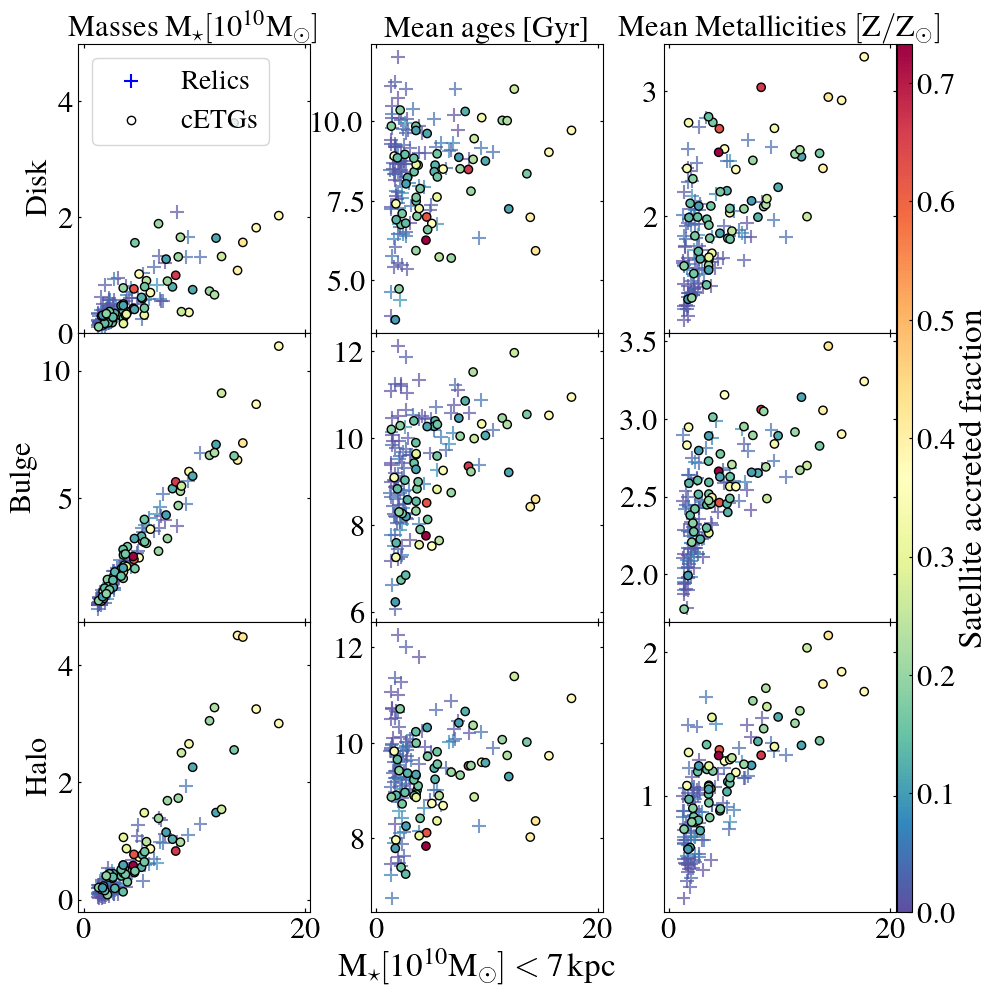}
\caption{Stellar total mass, mean age, and mean metallicity for each kinematical component (halo, bulge and disk) at $z=0$, color-coded by the satellite accretion fraction.} 
\label{Fig.5}
\end{figure}

 Relics and cETGs present non-negligible similar general trends at $z=2$ and dynamical disparities at $z=0$, considering the stellar mass dynamics over each epoch as shown in Fig.~\ref{Fig.3} and Fig.~\ref{Fig.4}. In Fig.~\ref{Fig.5}, it is possible to evaluate specifically the accretion effect through the stellar masses, mean stellar ages, and mean metallicities to Relics and cETGs at $z=0$. This figure is color-coded by satellite accretion fraction, where the low satellite accretion is represented by Relics subhalos displayed as blue `+' markers ($<10\%$ of accretion), while cETGs are scattered with more or less accretion along each frame to each kinematical component: halo, bulge, and disk.

Stellar age and metallicities were obtained in a similar manner as we obtained the stellar mass particles at each subcomponent, delimited by the orbital decomposition thresholds in the simulation. As result, the mean stellar ages obtained are $9.06$ Gyr, and $8.83$ Gyr respectively for Relics and cETGs. This stellar age is computed as the mean ages of each component together.
Concerning the mean metallicities of each component, the results obtained present a gradient of $\approx 0.3$ among the sample, with $Z/Z_{\odot} = 1.65$ and  $Z/Z_{\odot} = 1.99$ for Relics and cETGs subhalos respectively. Considering the cETGs extended accretion history compared to Relics, one expects a slight divergence between the mean ages and metallicity values for cETGs and Relics as shown. 

When examining the kinematic behavior of stellar mass, computed age, and metallicity distributions in relation to accretion fractions, it appears that the cETGs most important accretions do not seem to play a significant role in distinguishing their pathway to the Relic pathways without significant accretion. In the initial selection cut at $z=0$, we specifically selected massive, compact, and quiescent subhalos. These are then categorized based on their accretion fractions, utilizing a fixed percentage threshold to create distinct subsamples. Despite variations in their dynamical components at $z=0$ and slight imbalances in age and metallicity, the overall pathway does not reveal a clear distinction or a separate class of compact galaxies, even when the merger rate is substantial or as low as 10\%, as shown in the overall distribution overlap between the sample in Fig.~\ref{Fig.5}.

Relics and compact non-relic galaxies with similar stellar masses, sizes, integrated velocity dispersion, metallicities, and [Mg/Fe] abundances have been recently compared in \citet{2023Martin}, where they point out to a bottom-heavier IMF slope for relics and for non-relics, adding another constraint to this compact class of galaxies, separated by the star formation history.

\subsection{Environment}
\label{Sec:environment}
In this section, we present results concerning the local density and stripping correlations based on the global distribution of the sample in their environment.

\begin{figure}
\centering
\includegraphics[width=1.0\linewidth]{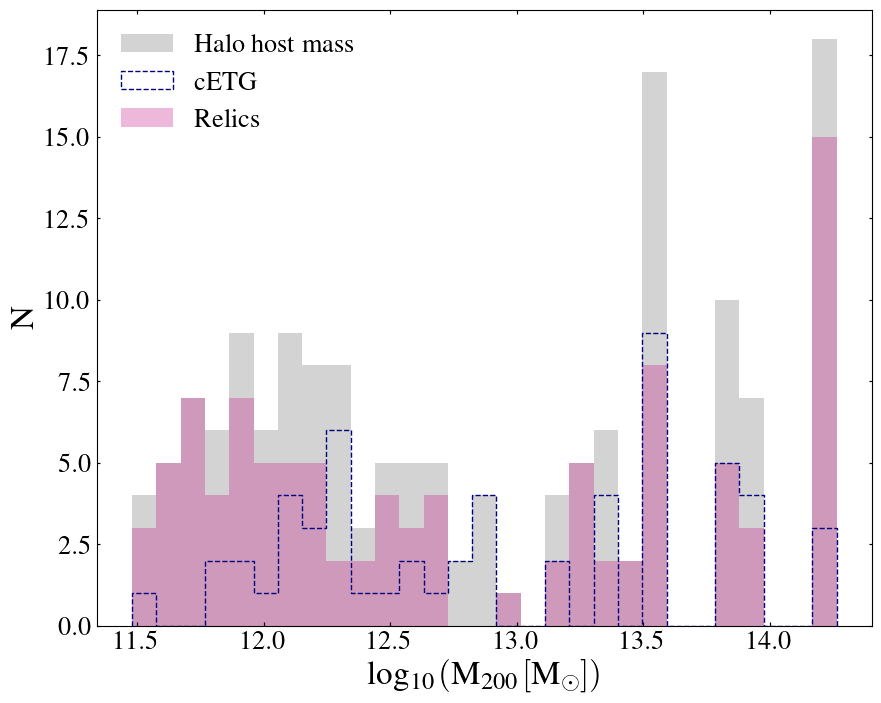}
\caption{Relics and cETGs distribution over $M_{200}$. The total sample distribution is displayed in gray on the background and the Relics and cETGs subsample are displayed in magenta and blue respectively, according to the legend.}
\label{Fig.6}
\end{figure}

\subsubsection{Global distribution}

TNG50 simulation is not suitable to explore environmental diversity, due to the focus of this size box on the high local resolution instead of great halo mass ranges. However, as we see in Fig.~\ref{Fig.6}, there is a substantial number of subhalos in halos with small masses (87 subhalos within $M_{200} \approx 10^{11.5} - 10^{13} \rm{M_\odot}$) and also in the larger halo mass (18 subhalos within $M_{200} \approx 10^{14} \rm{M_\odot}$), where such mass ranges can be compared to masses ranging small galaxy groups to a typical galaxy cluster. As result of the current sample distribution, one sees no clear relation concerning the halo masses (density proxy) preference for Relic or cETGs, where both groups are found in all ranges of the host-halo mass. The larger number of Relics in less massive halos may be attributed to their lower stellar masses compared to cETGs. As they have experienced fewer accretion episodes over time, this may also be reflected in their halo masses. Another possibility is the slight unbalance of the sample, composed of 99 Relics subhalos and 57 cETGs.

\begin{figure}
\centering
\includegraphics[width=1.0\linewidth]{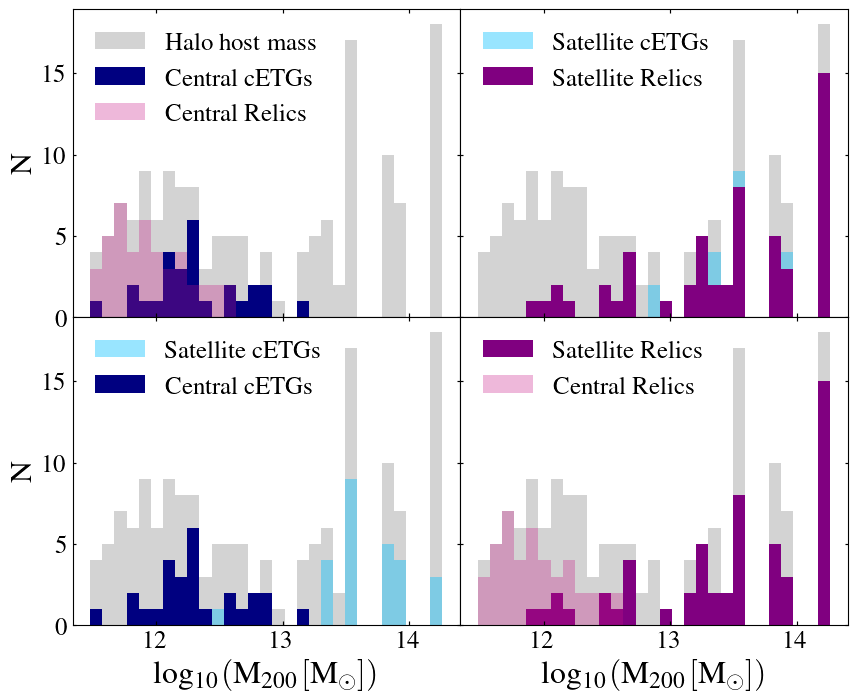}
\caption{Same as Fig.~\ref{Fig.6}, with frames separated by centrals and satellite distribution of each subsample.}
\label{Fig.7}
\end{figure}

Aiming to further refine the distribution of the subhalos in their environment, Fig.~\ref{Fig.7} shows the distribution of Satellites and Central galaxies among its host-halo masses for both samples: Relics and cETGs. The upper panels of Fig.~\ref{Fig.7} show only central subhalos (left), and satellite subhalos of the sample. Central subhalos are defined as the most massive galaxy of its host-halo, and satellites are defined as orbiting another massive subhalo in the same host-halo. Central subhalos are mainly located at small halo masses for both subsamples. Satellites are mainly located in massive halos, as expected due to the higher number of massive galaxies in higher densities. The lower panels indicate the central and satellite spread to cETGs and Relics separately. Fig.~\ref{Fig.6} and Fig.~\ref{Fig.7} set up the general distribution of the sample in their global environment, based on the masses of the host-halo at $z=0$.

The current distribution suggests the possibility of finding Relic galaxies as well cETGs on small galaxy groups masses, and as being part of a cluster of galaxies. This finding can be interpreted from two perspectives related to Relic galaxies: in less massive systems -- and therefore low-density environments, Relics are less likely to merge and increase the ex-situ fraction and therefore we expected to find this class of compact galaxies without significative accretion in this more isolated environment. On the other hand, as pointed by \citet{2016PeraltaArriba}, the associated high-velocity dispersions and the hot intracluster medium (ICM) can prevent the growth of an accreted stellar envelope through mergers. Another possibility in this cluster framework is to investigate the role of stripping, which can remove the envelope from the galaxy \citep{2009Kapferer,2022Peluso,2023Goller}.

\begin{figure}
\centering
\includegraphics[width=1.0\linewidth]{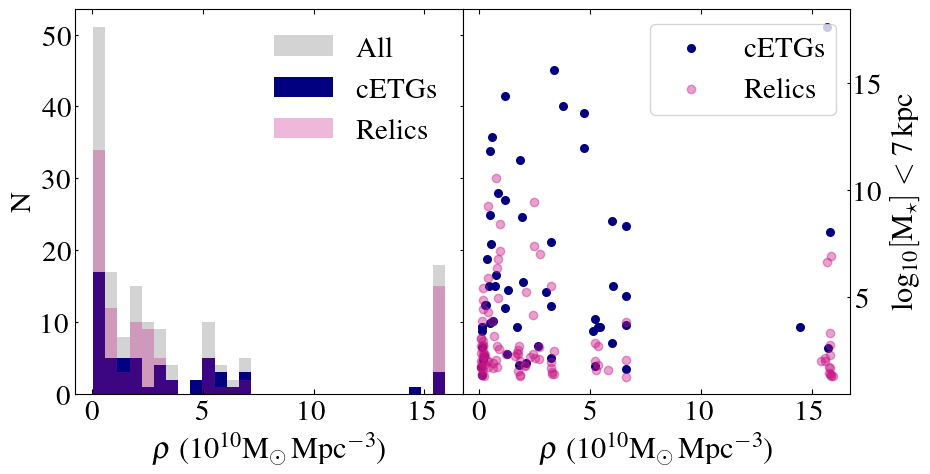}
\caption{Local density computed at the fixed aperture of 2 Mpcs around each galaxy. Left: The density sample distribution. Right: Subhalos total stellar mass dispersion depending on local density. Relics are displayed in magenta and cETGs in blue on both panels.}
\label{Fig.8}
\end{figure}

We quantified the global distribution density by calculating the local density using the aperture method described in \cite{2014mariacebrian,2016PeraltaArriba}, based on the total stellar mass around the galaxy within a fixed radius. Here we adopted a sphere of radius $R = 2$\,Mpc at $z=0$. The left panel of Fig.~\ref{Fig.8} shows the local density values for the sample, following the previous trends described before (see Fig.~\ref{Fig.7}). Considering the box size of TNG50, one cannot make strong assumptions related to the environmental preference of Relics. Meanwhile, it is possible to explore the total stellar mass relation with the environment density, as we see in the right panel of Fig.~\ref{Fig.8}. Although our selected Relics and cETGs are compact subhalos, the ones that had significant accretion are not found in high-density environments as compared to Relics, despite their higher masses. The larger number of Relics at high local density, compared to the cETGs selected at the same mass range, points to the preference of the Relics to a dense environment, following \citet{2016PeraltaArriba} results. In the current context, this may also point to a statistical bias, where due to the low number of cETGs compared to Relics (57 against 99), the number of cETGs in higher densities is not reliable. Another possibility is the absence of halo mass diversity, especially in the high-mass range. These two factors combined prevent us from making strong claims without further explorations with other cosmological simulations.

\begin{figure}
\centering
\includegraphics[width=1.0\linewidth]{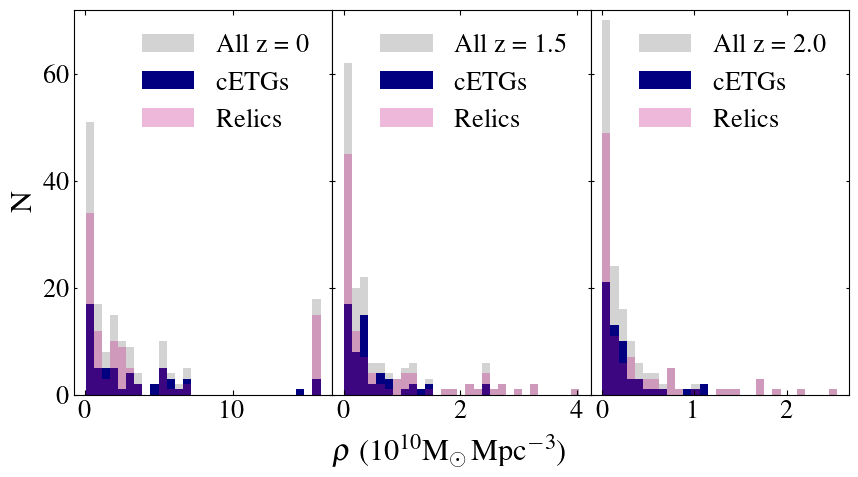}
\caption{Local density distribution in three different moments, at $z=0$ (left), $z=1.5$ (middle), and $z=2.0$ (right). The Relics and cETGs were traced using the progenitor of the subhalos at each given redshift.}
\label{Fig.9}
\end{figure}

We also investigate the environment of the subhalo progenitors of the sample. Fig.\,\ref{Fig.9} shows the evolution of the local density for cETGs and Relics from $z=2$, $z=1.5$, and $z=0$ (from right to left). Tracing back the local density of the subhalos progenitors, allows us to see the past environmental distribution until it leads to the current allocation. The larger Relics number (compared to cETGs) at higher densities since $z=2$, is remarkable because this group has not gone through significant accretion episodes, despite their environment density. The number density of subhalos in low-density environments changes slightly over time. As can be seen from Fig.\,\ref{Fig.7}, they are central and isolated, compared to the satellites in high-density environments. In Sect.~\ref{Sect.stripping} we discuss correlations between the features of subhalos located at low and high-density environments based on this spread.

\begin{figure}
\centering
\includegraphics[width=1.0\linewidth]{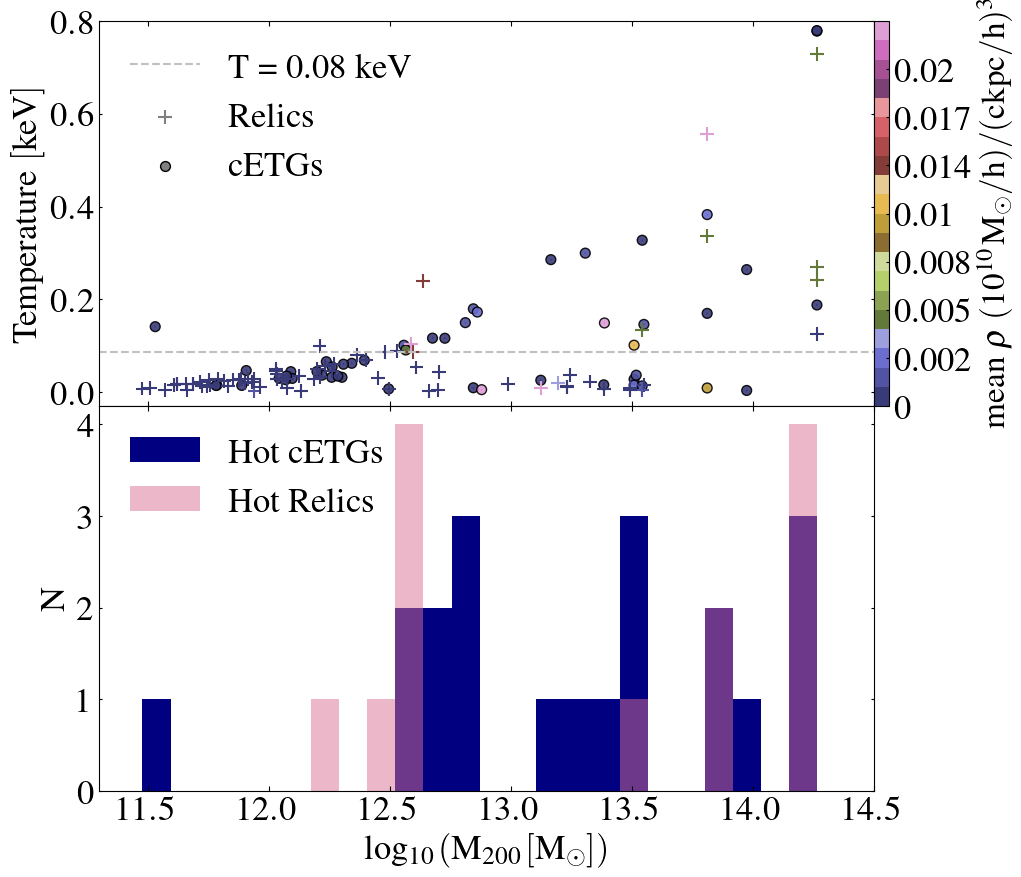}
\caption{Subhalos gas temperatures distribution. Upper panel: Gas temperature over $M_{200}$ color-coded by mean gas density of the subhalos with $M_{\rm{gas}}\,>0\,\rm{[M\odot]}$. The horizontal line represents the temperature of a fully ionized gas in keV. Lower panel: Hot Relics and cETGs distribution, defined as subhalos above the temperature limit on the upper panel.}
\label{Fig.10}
\end{figure}

Another perspective to gain insights into the relationship between the subhalos and their environment is by investigating the inner gas density and the associated temperatures. Some works reported the detection of hot X-ray emission atmospheres around two isolated Relic galaxies – Mrk 1216 and PGC 032873 \citep{2018ABuote,2018Werner}, suggesting temperatures in the order of $k_{B}\rm{T} \sim 0.6 - 0.7$ keV to the isolated Relics. Considering our setup, we investigate the gas temperature of the sample in Fig.\,\ref{Fig.10}. Hot subhalos are defined here as the ones with temperatures above $10^{6}\,\rm{K}$. Gas density and temperatures can be considered as a proxy to the X-ray emissivity (from \citealt{1995NFW}), as long as the temperatures are high enough to consider the fully ionized gas. This temperature limit is shown on the top panel of Fig.\,\ref{Fig.10}, selecting 125 subhalos with enough particles resolution to compute the temperatures, where 75 are Relics, and 50 are cETGs displayed on their host-halo masses, color-coded by the mean gas density. The large temperatures are found in higher density environments, as expected due to the influence of the surrounding intracluster medium to higher $M_{200}$ masses. Nevertheless, around $M_{200} \approx 10^{12.5} \rm{M_{\odot}}$, 33 subhalos have higher temperatures than the limit defined to select hot subhalos. This mass range is comparable to the Mrk 1260 mass, $M_{200} \approx 9.6 \times 10^{12} \rm{M_{\odot}}$ \citep{2018ABuote}. The subhalos with low temperatures (signed as $T \sim 0$ keV), contain a negligible amount of gas particles, preventing the temperature estimate, and should not be interpreted as low-temperature subhalos (to a more complete relation of the X-ray bubbles and temperatures estimative found on TNG50, see \citealt{2021Annalisaxray}). The hot subhalos global distribution is shown in the lower panel of Fig.\,\ref{Fig.10}, highlighting 33 hot subhalos, where 25 subhalos are located on low-density environment -- defined here as the subhalos below the $M_{200}$ mean masses of $10^{12.60}\,\rm{M_{\odot}}$, and 8 subhalos are placed in higher density cluster-like environment.

\begin{figure}
   \centering
\includegraphics[width=1.0\linewidth]{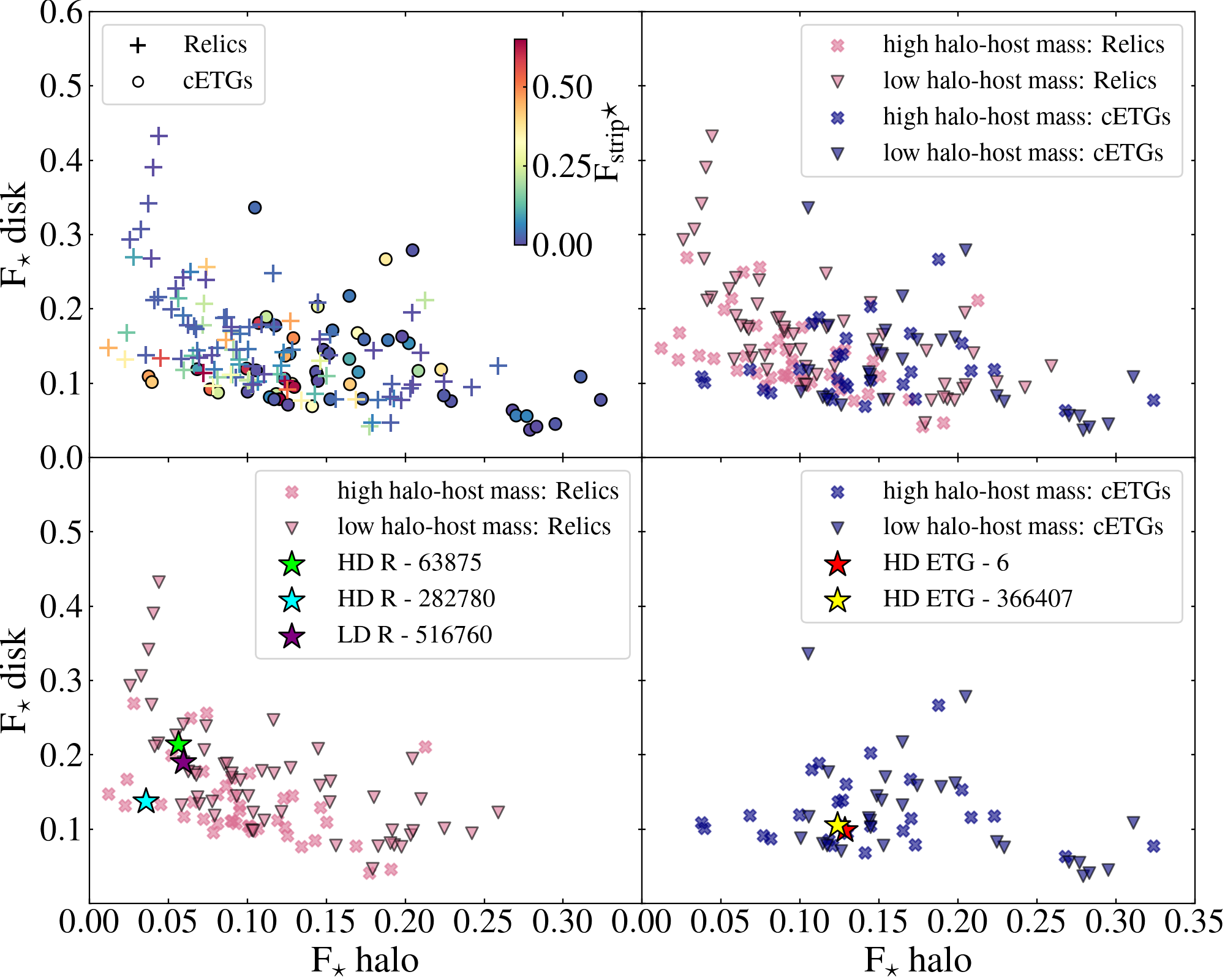}
\caption{Stellar $F_{\rm{disk}}$ over $F_{\rm{halo}}$ fractions of all subhalos separated by high and low host-halo masses. Upper left: all the Relics and cETGs distribution. Upper right: Relics in magenta with markers indicating low and high host-halo masses, as well cETGs in blue, according to the legend. Lower left: Same plot only with Relics. Lower right: Same plot only with cETGs. The colored stars represent the 5 Relic subhalos from \citet{Floresfreitas2022}, where HD R is high-density environment Relics, and LD R is low-density environment Relics. Analogously, HD ETG means high-density environment to the ETG.} 
\label{Fig.11}
\end{figure}

\subsubsection{The stripping correlations}
\label{Sect.stripping}

Besides the accretion, we explore the stripping history of the sample related to their environment. The tidal forces between the galaxy and the environment disturb the galaxy's structure, interfering in the dynamics and morphology, or even leading to their disruptions (e.g.~\citealt{1983AMerritt,2022Pallero}). Fig.~\ref{Fig.11} (upper-left panel) quantifies the stellar mass stripped fraction from the $F_{\rm{disk},{\star}}/F_{\rm{halo},\star}$ relation. The stellar mass stripping is defined as the fraction between the stellar mass at $z=0$ over the maximum stellar mass through their history: $F_{\rm{strip}} = 1 - (\rm{M_{\star}}_{(\rm{z = 0})}/\rm{M_{\star}}_{(\rm{max})})$. Most of the subhalos (110) have less than 15\% of stellar mass stripped (73 Relics and 37 cETGs), while 23 subhalos have $>30\%$ of the stellar mass stripped, where 13 are Relics and 10 are cETGs. It is possible to notice from Fig.~\ref{Fig.11} the distinct kinematic component dependence for Relics with larger stellar disk fraction ($>0.30$), and small stellar halo fraction ($<0.05$), compared to the cETGs with larger stellar halo fraction ($>0.25$), and lower stellar disk fraction ($<0.1$). They stand out to opposite sides of the stellar disk-halo relation, despite both groups showing a low stripping fraction. This suggests that cETGs, which have significant accretion history differ from the Relics in the extreme cases of stellar disk and halo fractions. For subhalos with high stripping fraction (i.e. $>15\%$) we see no clear correlation with each component, disk, and or halo fraction.

\begin{figure}
    \centering
\includegraphics[width=1.0\linewidth]{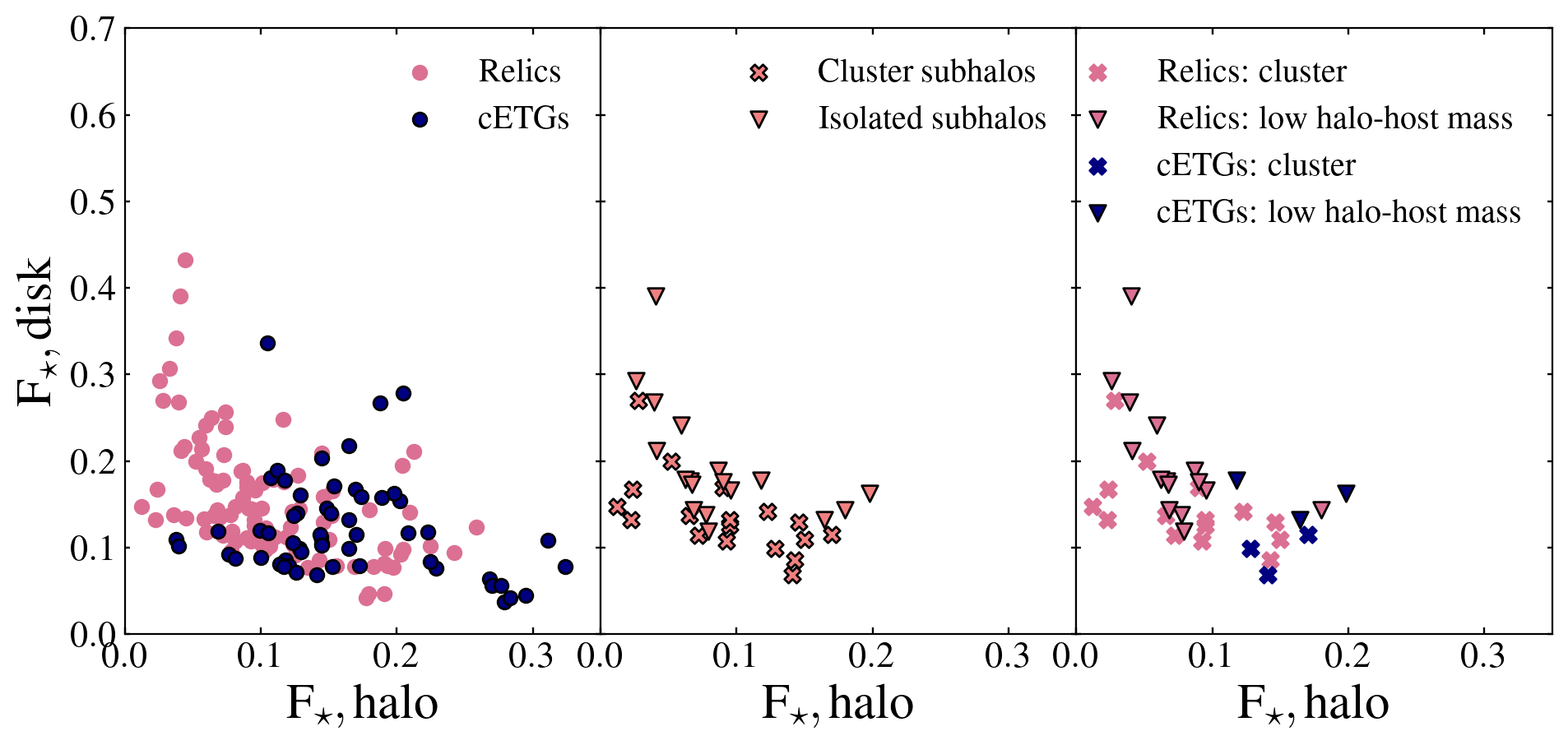}
\caption{Stellar $F_{\rm{disk}}$ over $F_{\rm{halo}}$ fractions to subhalos isolated and placed in the cluster. Left: All the distribution of the sample. Middle: Cluster subhalos and isolated subhalos distribution separated by different markers. Right: Same as middle with Relics and cETGs identified and displayed by different colors.}.
\label{Fig.12} 
\end{figure}

\begin{figure}
\centering
\includegraphics[width=\linewidth]{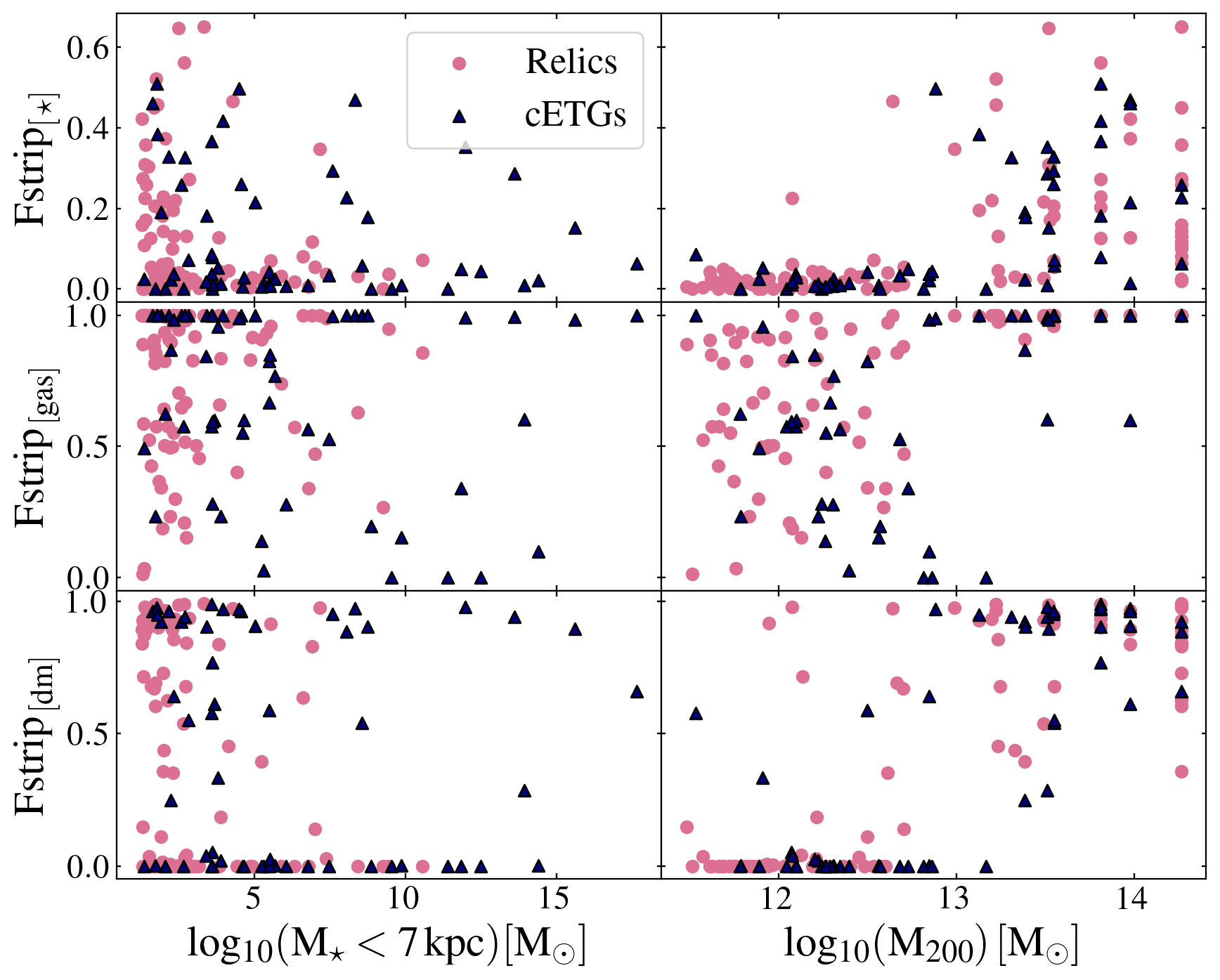}
\caption{Stripped fraction of the components: stars, gas, and dark matter (DM) over the total stellar mass of the galaxies (left) and $M_{200}$ (right) at $z=0$.}
\label{Fig.13} 
\end{figure}

The stripping correlation with the environment has been widely investigated through numerical simulations (e.g.~\citealt{2013Chang,2021Engler,2022MonteroD}). Here we associate the stripping correlations with the environment also in Fig.~\ref{Fig.11} (Upper-right panel). We define the mean mass of the host-halo $M_{200} = 10^{12.60}\,\rm{M_{\odot}}$, as the mass limit to separate high and low-dense environments. Subhalos below the mean value are defined as being part of the low host-halo mass regime, and in comparison, subhalos above the mean value are defined as being part of the high host-halo mass. This definition is convenient since it reduces the gap between the number of subhalos in low-density environments compared to subhalos in the cluster environment of the simulation and keeps the local density proportion.

The upper-left panel of Fig.~\ref{Fig.11} shows the stripping fractions over the sample, and the upper-right panel shows the same distribution with markers indicating the environment masses for both groups according to the legend. Considering the TNG50 framework, the high-density galaxy's location matches with the high-halo masses ranges, once one can use the high-density environment to refer to the considered high host-halo masses.

The upper-right panel shows Relics with low stellar stripping that have high stellar mass disk fraction ($>0.3$) and low stellar mass halo fraction ($<0.05$) are located in a low halo mass environment. On the other hand, cETGs on the opposite side compared to Relics with similar stripping fractions are mixed in their environment. Two of eight subhalos are found in high halo masses, and the remaining are located in a low halo mass regime. The lower-left panel shows a general distribution for the Relics separated by their environment, evidencing that not all the Relic subhalos in the low-density environment will show a stellar disk dominance over the halo, even with low stripping fraction associated. Relic galaxies that evolve in the low-density environment are more likely to keep their disk, but not in all cases. Concerning the cETGs on the lower-right panel, the non-environmental correlation seems even more evident.

We also investigate these (kinematical) morphological features considering the subhalos located in the most massive cluster of the simulation, named `Halo 0' with $M_{200} = 2 \times 10^{14} \rm{M_{\odot}}$, to compare with the less massive host-halos of the sample -- defined here as the isolated subhalos, with halo masses between $M_{200} = 10^{11.47}~\rm{to}~10^{11.81} \rm{M_{\odot}}$. The aim is to certify the consistency of the distribution shown in Fig.~\ref{Fig.11}, adopting the extreme cases. In other words, we explore the $F_{\rm{disk},{\star}}/F_{\rm{halo},\star}$ dependence on the environment by looking at the subhalos located in a cluster-like halo, to compare with the ones located in the field/isolated. There are 18 subhalos of the sample located in the cluster, which we compare with another 18 subhalos of the sample located in the field. The equal subhalo's number, separated only by the opposite environment characteristics, allows us to certify if the tendencies found before are consistent considering the field and a cluster scenario. The results are shown in Fig.\,\ref{Fig.12}. The right panel of the figure identifies the class of the subhalos and shows their location. The central panel shows the 18 subhalos selected in the cluster among the 18 subhalos selected in a field environment, without distinction between the samples (Relics and cETGs). Here we can see that the cluster subhalos overlap with the isolated subhalos indicating no discernible trend. The right panel identifies the group of the selected subhalos. This result supports the trends aforementioned, that there is not a clear correlation between the Relics and cETGs distribution in their environment, considering the stellar disk and halo fraction, even in extreme cases.

From an overall stripping framework, Fig.~\ref{Fig.13} shows the stripped stars, gas, and dark matter fractions as a function of the total stellar mass of the galaxies within 7 kpc (left panel) and the halo host mass (right panel). Considering the stripped fractions and the total stellar mass it is possible to notice that the stars and dark matter are more bound to the galaxies, while the gas particles which are susceptible to hydrodynamic transformations, can be easily depleted from them \citep{2012Zemp}, and as a result, almost all subhalos display low amounts of gas mass available. Despite the non-collisional treatment, the dark matter is also depleted in a significant number of subhalos Relics and cETGs,  more than stars. This can indicate that these subhalos with low dark matter and gas fractions passed through disruptive events that led to the increase of the stripped fractions. However, this does not affect the stellar content of these galaxies. Such trends appear to be independent of the type of galaxy (cETG or Relic). Concerning the dark matter content of Relic galaxies, \citet{2023Comeron} has investigated the dark matter fractions of the Relic galaxy NGC 1277 (D = 71 Mpcs; \citealt{Yildrim2017}) located in the dense environment of the Perseus Cluster \citep{2014Trujillo}. They found a negligible dark matter fraction within $5\,R_{e}$, suggesting that is possible for some Relic galaxies to be depleted dark matter due to the stripping processes as they interact with the surroundings in a galaxy cluster. Considering the current framework, we found 15 subhalos within $M_{200} \approx 10^{14}\rm{M_{\odot}}$ sharing physical similarities with NGC 1277, in terms of lack of dark matter ($f_{\rm{strip}}>60\%$), and compactness (with $\rm{R_{e}}<4$ kpc).

Another approach is to correlate the stripping fraction with the environment. From this point of view, we can compare the stripping fraction to galaxies in low, and high-density environments. Here we define a low-mass density environment as halo masses lower than the mean density, as defined before. The right panel of the figure (Fig.~\ref{Fig.13}) shows that in proportion, the galaxies in low-mass regimes keep their stars and DM in comparison with galaxies in high-density environments. Whereas in high-mass environments the galaxies (both ETGs and Relics) are highly DM-stripped. The galaxies in high-density environments lose almost all their gas mass while even the galaxies in low density have their gas stripped. It is noteworthy that AGN effects such as gas removal are not considered in this work, although cannot be completely excluded due to our stripping definition. Further explorations about the nature of the gas absence and other components in low-density environments are needed in future works.

The environmental framework sheds light to interpret the stripping shown on the galaxy set, where the highest stripped fraction is due to the place where the galaxies are. Based on the subhalos distribution shown in Fig.~\ref{Fig.7}, and in the stripping correlation with the host-halo mass, it is clear that the environment plays a more important role than the mass of the galaxy in defining the stripping, independent of the group of the galaxies or the class (central or satellite) in low host-halo mass regime. Satellites in low density (therefore, more isolated) have lower stripped fractions as much as the central galaxies in this regime. On the other hand, in the high-mass regime of the host-halos, stripping is dominant to all components.

\section{Summary and discussion}
\label{Sec:summary}
Relic galaxies act as fossils of an ancient Universe, becoming for this reason the ideal candidates to investigate in situ star formation, stellar kinematics, gas and dark matter fractions, and other preserved features through time. From an observational point of view, investigating this class of frozen galaxies in detail can provide us clues to understand the evolutionary processes of massive and compact galaxies, and from a theoretical point of view, they can act as the key to improving models of synthesis of stellar population and IMF constraints. In this work throughout TNG50-1, we investigate the assembly history of 156 old compact and massive early-type subhalos separated by their accretion history, where 99 subhalos had less than $10\%$ of accretion through time (Relics) and 57 had extended accretion history (cETGs). We explore the local internal galactic dynamics and their global environmental distribution using the TNG50 cosmological simulation. We employ the orbital decomposition method to investigate the stellar dynamics from the disk, bulge, and halo kinematical components over $z=0$, $z=1.5$, and $z=2$. The main results are summarized as follows:

i) Overall, the total sample composed by cETGs and Relic galaxies share similarities in terms of assembly history, considering general aspects such as age, metallicity, compactness, and environment. At $z=2$, our sample exhibits comparable dynamic characteristics in all kinematic components (bulge, disk, and hot inner stellar halo), and as time progresses (about $z=1.5$), the stellar dynamics of each component undergoes changes, due to mergers, creating a stellar dynamic distinction between cETGs and Relics at $z=0$ for all listed kinematic components, as well as for the $\rm{F_{disk}~ and~F_{halo}}$ fractions. Despite the difference in stellar dynamics of these components at $z=0$, these galaxies do not follow a different evolutionary path. Thus, mergers (regardless of minor or major) do not act as an essential factor in transforming galaxies that were compact at $z=2$ into full massive ETGs at $z=0$.

ii) To further constrain the scenario proposed above, we examine the spread of stellar ages and metallicities among the stellar masses across the components (Fig.~\ref{Fig.5}) to better illustrate the role of accretion over disk, bulge, and halo. It is expected that some cETGs are outliers in the distribution for higher masses ($\rm{M}_\star$ within 7 kpc) due to late accretion. Except for these outliers, the subhalos that experienced significant accretion (50\%, 60\%, and 70\%) overlap with Relic subhalos in each evaluated component, following a consistent dynamic pathway leading to a slightly younger (8.83 Gyr compared to 9.06 Gyr) and more metal-rich ($Z/Z_\odot = 1.99$ compared to $Z/Z_\odot = 1.65$) compact group, contrasting with the Relic group, both observed at $z=0$.

iii) The overall environmental spread of the sample reveals Relics alongside cETGs covering all ranges of halo masses in the explored volume. Some works associate distinct pathways of compact elliptical galaxies (cEs) in isolated environments compared to those compact galaxies being satellites of a host-associates galaxy \citep{Ferre-Mateu2017,Kim_2020,Deeley2023}. In Fig.~\ref{Fig.7} we show that both subsamples are composed by central subhalos (without an associated host) and satellite subhalos (with an associated subhalo host). Considering the previous kinematic exploration without environmental distinction between centrals and satellites, we can suppose that the environment or the presence of association does not impact internal dynamic processes. In other words, in this current context, the environment is not the decisive factor for distinguishing the pathways of massive central and satellite cETGs. As pointed out in Sect.\ref{Sec:environment}, TNG50 can introduce a bias due to its volume, where there are more galaxy group-like masses compared to cluster-like diverse environments. Nevertheless, this bias is mitigated when we analyze the extreme of the distributions in Fig.~\ref{Fig.12}.

iv) In addition to the overall distribution at $z=0$, we investigated the evolution of local density (within 2 Mpc) for progenitors at $z=2$ and $z=1.5$ of current subhalos. Our findings indicate a higher number of Relics in denser environments compared to cETGs in the same mass range at all evaluated redshifts. As previously mentioned, such trend aligns with the results from \citet{2016PeraltaArriba}, where the possibility of environmental effects of the ICM can inhibit the accretion, hindering their growth and maintaining compactness since then. In a low-density environment, it is plausible to consider that the growth is impaired due to a lack of sufficient surrounding material. Concerning the ongoing growth of compact galaxies in isolation, \citet{Deeley2023} emphasizes that during the initial stages of cE evolution, gas gradually becomes concentrated at the center. Meanwhile, the compressed star formation process continues to accumulate stellar mass within these core regions, leading to an inner increase concentration of the galaxy. While agreement has been found regarding the Relics placed at high-density environments, further explorations are needed to constrain the parameters space considering other simulation volumes, especially the ones that provide a variety of dense environments.

v) We also linked the general aspects of the sample distribution with the subhalos gas temperature. We defined hot subhalos as those having gas temperatures greater than $10^6$\,K. Only 125 subhalos have sufficient gas particles to compute temperature and density, among which 33 subhalos were classified as hot subhalos, including both cETGs and Relics (Fig.~\ref{Fig.10}). The subhalos' temperature increases with the halo mass in $M_{200}$, with the exception of 1 subhalo having high temperature despite being in a low-density environment; while an exception, it is not uncommon as there are observations of hot Relic galaxies that are not in cluster environments \citep{2018Werner,2018ABuote}.

vi) We quantified the fraction of stellar stripping across the sample and correlated it with the environment. Subhalos in low-density environments were defined as those below the average halo masses in $M_{200}$, whereas subhalos in high-density environments are defined above this average. The overall distribution of subhalos is described in terms of $\rm{F_{\star disk}}$ and $\rm{F_{\star halo}}$, classified according to their respective environments in Figure~\ref{Fig.11}. Naturally, higher rates of stellar stripping are correlated with higher mass environments. However, the results for low-density and low-stripping cases require further attention. Within this context, some Relic subhalos maintain their disks (Fdisk > 0.3 and Fhalo < 0.05). In the same way, in low-density environments with low stripping fraction, some cETGs show smaller disks despite displaying a prominent hot inner stellar halo (Fdisk < 0.1 and Fhalo > 0.25) tracing the ancient mergers of this group \citep{2022ZhuLing}. In such scenario, due to environmental conditions, Relics that have not undergone significant accretion or stripping are found in isolation, with their disk structure preserved. In another context within the same low-density setting, cETGs with extended star formation history due to accretion exhibit an absence of a distinct disk structure. Further investigation is needed to enhance the statistical significance of these findings, both in terms of sample size and diversity of environments. Regarding the high stripping fractions and isolated disk-like Relics, our exploratory results align with the findings by \citet{Deeley2023}, where they performed observations combined with TNG50 simulations of compact ellipticals (cEs) indicating two primary pathways for cEs' formation: through the stripping of a spiral galaxy associated with a host galaxy, and through the gradual growth of stellar mass in isolated environments.

vii) Additionally to the stellar stripping, in Fig.~\ref{Fig.13} we investigated the gas and dark matter stripping as a function of the local dependence of the total stellar mass (within 7 kpc) and the global dependence considering their halo in $M_{200}$. The stellar component is shown to be the most bound component considering both perspectives and in contrast, the gas mass is the most volatile component. It is reasonable for the gas particles to be easily depleted due to the hydrodynamic interactions compared to the collisionless components (stars and DM), and even in lower $M_{200}$ masses, the gas stripping is highly notable. It is noteworthy that the stripping fraction defined here does not exclude AGN effects or other mechanisms besides the environmental stripping due to infall for example. In some cases, gas and dark matter are almost completely depleted of the subhalos, although DM particles do not undergo collisional effects. The dark matter deficiency of NGC 1277 in its inner regions and within $6\,\rm{kpc}$ has been reported in the literature \citep{Yildrim2017,2018Beasley,2023Comeron}. The dark matter stripping could be due to an interaction with the galaxy cluster at early times, or due to a dark matter deficiency \textit{ab initio}. The second scenario is supported by dynamical modeling and the lack of a blue GC population \citep{2018Beasley}. Nevertheless, we do not fully understand how stripping affects GC populations. One finds bimodal optical distributions over the full galactocentric radii, but blue GCs in NGC 1277 seem to be an exception to this. Further investigations concerning the cases with a massive concentrated amount of stars with low DM fractions in dense environments are required. High-resolution cosmological simulations alongside high-performance observation data of radial velocities of GCs out to large galactocentric radii are of utmost importance to be obtained. The latter can be achieved with JWST-NIRSpec.

In previous work, \citet{Floresfreitas2022} found five Relic-analogs at $z=0$ using the TNG50 simulation and comparing it with observational constraints. As mentioned in Sect.~\ref{Sec:Methods and data}, the criteria to select Relic candidates vary from less to more restrictive parameters. Here we applied the mass, size, sSFR, and age criteria alongside the accretion fraction of Relics to compare with solely compact galaxies with the same criteria but showing extended star formation history. As a matter of discussion, Fig.~\ref{Fig.11} shows Relics from Flores-Freita's work as colored stars and their environmental classification. Subhalos 63875, 282780, and 516760 matched with our Relic definition, while subhalos 6 and 366407 are defined as compact ETGs due to the applied sample selection criteria ($\rm{F_{accretion}}<10\%$), although all of them are the best Relic models based on the observations. We also explore the possibility of an accretion rate varying according to the stellar mass, assuming a more restrictive accretion rate (5\%) for the low-mass regime  ($10^{10} - 10^{10.5}\,\rm{M_\odot}$). The results were similar to those obtained with a fixed accretion fraction of 10\% for the entire sample.

In conclusion, we performed a theoretical dynamical analysis using the orbital decomposition method applied to the TNG50 cosmological simulation to investigate a sample of 156 massive, compact, and quiescent simulated galaxies at $z=0$, $z=1.5$, and $z=2$. The employed orbital decomposition method can be applied to massive compact ETGs by constructing orbit-superposition models to IFU data, providing a way to distinguish the different dynamics features of compact ETGs at $z=0$, recovering their dynamical past from the stellar kinematics. This method applied in a high-resolution simulation allows us to understand finer dynamic processes involving compact galaxies through time, enabling a new interpretation of observational data, and/or refinement of theoretical models for these objects at high redshifts. We found that compact galaxies with significant accretion exhibit non-negligible similarities in the evolutionary dynamic pathways to those compact galaxies without significant accretion. Although the evolutionary path does not lead to different branches, the stellar dynamics at $z=0$ are significative different, indicating that mergers impact the hot inner stellar halo regions of galaxies with extended accretion history. The reported comparable pathways between cETGs and Relics persist even when exploring extreme environmental cases, suggesting that the effects of local galactic dynamics on these objects outweigh the influences from the environment.

\section*{Acknowledgements}
This work is funded by Coordenação de Aperfeiçoamento de Pessoal de Nível Superior (CAPES Proj. 0001) and the Programa de Pós-Graduação em Física (PPGFis) at UFRGS.
MTM acknowledges the Brazilian agency Conselho Nacional de Desenvolvimento Científico e Tecnológico (CNPq) through grant 140900/2021-7. ACS acknowledges funding from the CNPq and the Rio Grande do Sul Research Foundation (FAPERGS) through grant  CNPq-314301/2021-6 and FAPERGS/CAPES 19/2551-0000696-9. CF ackwoledges funding from CNPq (433615/2018-4 and 314672/2020-6) and FAPERGS (21/2551-0002025-3). MC acknowledges funding from CNPq/PIBIC. This work is supported by the CAS PIFI programme 2021VMC0005. We thank Rodrigo Flores-Freitas for the useful insights from the previous work on Relics candidates on TNG50 simulation.

\section*{Data Availability}
The TNG50-1 run from IllustrisTNG simulations project is publicly available at \url{www.tng-project.org/data} \citep{2019Nelson}. Additional data from this work are available on reasonable request from the corresponding author.
 


\bibliographystyle{mnras}
\bibliography{relics} 







\bsp	
\label{lastpage}
\end{document}